# Information Needs and Practices Supported by ChatGPT

**Gorichanaz, Tim**  Drexel University, United States | gorichanaz@drexel.edu

## ABSTRACT
This study considers ChatGPT as an information source, investigating the information needs that people come to ChatGPT with and the information practices that ChatGPT supports, through a qualitative content analysis of 205 user vignettes. The findings show that ChatGPT is used in a range of life domains (home/family, work, leisure, etc.) and for a range of human needs (writing/editing, learning, simple programming tasks, etc.), constituting the information needs that people use ChatGPT to address. Related to these information needs, the findings show six categories of information practices that ChatGPT supports: Writing, Deciding, Identifying, Ideating, Talking, and Critiquing. This work suggests that, in the AI age, information need should be conceptualized not just as a matter of "getting questions answered" or even "making sense," but as skillfully coping in the world, a notion that includes both understanding and action. This study leads to numerous opportunities for future work at the junction of generative AI and information needs, seeking, use and experience.

## INTRODUCTION
Since the public release of ChatGPT in November 2022, generative AI products have been influencing the ways people need, seek and use information. Almost immediately, ChatGPT was branded a "Google killer," suggesting a sea change to the way people find information online (Kleinman, 2023). Since then, a slew of other generative AI products have emerged in this competitive landscape. While many products are styled as chatbots or assistants, some explicitly invoke information and search. For example, Perplexity AI dubs itself "the modern search engine." Generative AI tools represent a small but growing market share in online search (Dodge, 2024); in January 2025, it was reported that Google's global search market share was below 90% for three consecutive months, a first since 2015 (Goodwin, 2025).

The field of information behavior (broadly defined) has a long history of examining how human information engagement changes as new technologies come out. To date, however, there has not yet been much research on generative AI in the field. Most of the research on how and why tools such as ChatGPT are used is limited to psychological motivation and user experience.

This paper presents a study of how and why ChatGPT, framed as an information source, is used. In particular, this study examines people's motivations and use cases for turning to ChatGPT to get information. Linking generative AI use to information needs and information practices/behavior in this way sheds further light on why people currently use generative AI products. Moreover, it has the potential to enrich the existing theory within information research.

This study responds to the following research question: *What information needs and practices does ChatGPT support?* While many generative AI products exist, this study focuses on ChatGPT because it currently has the largest market share in the generative AI landscape (Moore & Zhao, 2025) and has seen the most academic study to date. To frame this study, the paper begins by reviewing the literature on information needs and then on the motivations of and goals for using ChatGPT.

## BACKGROUND

### Understanding Information Need
This study examines why people use ChatGPT, understood as an information source. As such, it looks into people's information behavior and practices with ChatGPT with the key question of what information needs are being addressed by ChatGPT.

Information need is a foundational concept in the field of information behavior and practices, which encompasses the myriad ways information influences human activity (and vice versa). Defined roughly, an information need is what causes, drives or motivates information seeking. However, the concept has been troubled for decades; problems include distinguishing needs from wants, defining whether needs must be conscious or could be unrecognized, determining whether information needs can be accounted for by other human needs (e.g., hunger), etc. These issues are reviewed and historically contextualized by Given et al. (2023, pp. 26–32).

In their literature review, Case and Given (2016, ch. 5) suggest that approaches to information need range on a continuum from objective to subjective. "Objective" conceptualizations tend to focus on rational, pragmatic drivers of information for clear and concrete instrumental purposes. Scholars taking this perspective tend to be oriented toward the design of systems for information retrieval. At the other end of the continuum, scholars taking the "subjective" conceptualization focus on the feelings that drive information seeking, such as anxiety or unease, setting aside rational goals. Case and Given highlight three major ways of conceptualizing information need in the literature: as answering a question, as reducing uncertainty, and as making sense.

1. Information need **as answering a question** was developed by Taylor (1968), which continues to be the most frequently cited work on information needs. Taylor described four levels or stages of question formation when a person engages with an information system: first, a visceral need is a somewhat ineffable, fuzzy sense of dissatisfaction; then a conscious need is that dissatisfaction roughly defined mentally; then a formalized need is that mental description stated in linguistic terms; and finally a compromised need is that linguistic description as presented to an information system.

2. Information need **as reducing uncertainty** was most prominently developed by Belkin and colleagues in the "anomalous state of knowledge" (ASK) framework. This understanding of information need focuses in particular on the visceral need notion of Taylor, suggesting that a person's recognized gap or uncertainty in their understanding is a key driver of using an information retrieval system (Belkin et al., 1982). On this framework, an information seeker is constantly checking back to their ASK to determine if it has been resolved.

3. Lastly, information need **as making sense** is communicated in the works of Dervin (1983), who conceptualizes information need as the result of a problematic situation in which a person finds themselves; the information need is how the person makes sense of their situation in effort to resolve its problems.

These broad approaches to conceptualizing information need are not necessarily incompatible. They emphasize different components of human situations. Together, they help clarify what information need means. However, the works mentioned above are general in nature and do not provide much detail on, for example, the different types of information needs that people may experience, which may be helpful in efforts to better understand and address information needs. Notable to that end is the work of a 1974 technical report prepared for the U.S. Education Department (then the Office of Education), which provides a method for describing and categorizing information needs along four dimensions: what user behavior the information is meant to support; the nature, amount and source of information being sought; the quality of information being sought; and the timeliness of information needed (Faibisoff & Ely, 1974).

This framing emphasizes that information needs are closely tied to people's information behavior and practices. Indeed, the two are somewhat circularly defined, and indeed an information need cannot be observed directly but can only be inferred through looking at how people interact with a system or otherwise engage with information (Case & Given, 2016).

**From Information Need to Information Behavior and Practices**

Like information need, there have been several decades of theory development in information behavior (and more recently, information practices). Theories in information behavior often illustrate the process a person goes through in a given situation of information needing, seeking and/or using (see Fisher et al., 2005). Wilson's (1999) general model of information behavior and Kuhlthau's (1991) model of the information search process are illustrative examples. Both show the stages a person passes through on the road to becoming informed; Wilson's emphasizes activities of a more objective nature, while Kuhlthau's is more situated in the person's subjectivity.

Another approach to theory presents a typology of the kinds of activities, practices or behaviors people undertake in a given situation, not necessarily followed in sequence. Hektor's (2001) general classification of information activities is a good example. This typology describes eight different high-level activities a person may do with information, such as "search and retrieve" (finding information), "monitor" (returning to familiar sources), and "dress" (creating information).

This style of theoretical contribution is particularly common in the literature on information practices, which focuses on the repeated or habitual activities that people engage in within an information environment. Caidi et al. (2024) provide a recent typical example in their study of employment seeking among STEM-trained immigrant women in Canada. Though not a strict typology, their research presents a narrative of different information practices in this context, such as navigating dissonance and negotiating alternatives. Other works do present a tighter typology, such as Potnis and Winberry's (2022) delineation of seven information practices that people use to overcome information vulnerability, mapped on three axes; these include practices such as selectively choosing information sources, building a support system, and adopting communication rules and norms.

**Motivations and Goals for Using ChatGPT**

Stated plainly, the central question in this study is: *Why do people use ChatGPT?* A "why" question can be answered in multiple ways. Aristotle categorized these ways into a framework of four "causes": material (what something is physically made of), formal (its structure), efficient (the impetus for it), and final (the goal or purpose) (*Metaphysics*, V.2). Most relevant to the present study are the efficient causes (i.e., what motivates people to use ChatGPT) and final causes (what people are trying to accomplish by way of ChatGPT). The material and formal causes would describe the technical details of ChatGPT and the sociotechnical environment in which ChatGPT is used.

To begin with people's motivations for using ChatGPT, previous studies have approached this question through a range of established theoretical models:

- the classic Technology Acceptance Model (TAM), which describes the factors leading to people's acceptance and use of a technology, such as perceived usefulness and perceived ease-of-use (Kim, Kim, et al., 2024)

- the Unified Theory of Acceptance and Use of Technology (UTAUT), which attempts to consolidate previous models, including TAM (Li, 2025; Menon & Shilpa, 2023)

- UTAUT2, an updated version of UTAUT incorporating aspects such as hedonic motivation and perceived price value (Huy et al., 2024; Yin et al., 2023)

- the AI Device Use Acceptance model (AIDUA), which adapts the above models specifically for AI technologies, recognizing for instance that with AI acceptance and rejection may co-exist (Ma & Huo, 2023)

- Task–technology Fit (TTF), a model that predicts how well technology helps users perform tasks (Huy et al., 2024)

In addition, a number of other frameworks have been used to help guide research questions and organize findings regarding the uses and motivations for use of ChatGPT: Uses and Gratifications (Baek & Kim, 2023; Boryung & Brenton, 2024; Kim, Kim, et al., 2024; Skjuve et al., 2024); Stimulus-Organism-Response (Pham et al., 2024); Cognition–Affect–Conation (Zhou & Zhang, 2024); Learning–Using–Assessing (Sun et al., 2024); and the Pragmatic/Hedonic model of user experience (Skjuve et al., 2023). For the most part this work has employed questionnaires, but three studies used interviews (Menon & Shilpa, 2023; Sun et al., 2024; Wolf & Maier, 2024), one used a focus group (Jung et al., 2024), one a diary study (Kobiella et al., 2024) and secondary data analysis (Kim, Lee, et al., 2024).

Despite the range of theoretical approaches in prior work and some variance in findings, some patterns are evident. According to the current literature, major contributors to ChatGPT use include:

- expectations of its performance, usefulness and efficiency for completing a task (Boryung & Brenton, 2024; Kobiella et al. 2024; Li, 2025; Ma & Huo, 2023; Menon & Shilpa, 2023; Skjuve et al., 2023; Sun et al., 2024; Wolf & Maier, 2024; Yin et al., 2023)

- expectations regarding its ease-of-use and the effort necessary to work with it (Li, 2025; Ma & Huo, 2023; Menon & Shilpa, 2023; Skjuve et al., 2023; Sun et al., 2024; Wolf & Maier, 2024)

- social influence and word of mouth from other users (Boryung & Brenton, 2024; Huy et al., 2024; Li, 2025; Ma & Huo, 2023; Menon & Shilpa, 2023; Quan et al., 2024; Yin et al., 2023). Similarly, using ChatGPT leads to the intention to continue using it and recommending it to others (Huy et al., 2024; Kim & Baek, 2024; Quan et al., 2024).

Other contributors to ChatGPT use have seen more scattered support. For example, the idea that ChatGPT is superior to alternatives may motivate some users to choose ChatGPT (Skjuve et al., 2023), though Kim and Baek (2024) found that quality of alternatives did not affect users' commitment to ChatGPT.

Other studies have looked into more specific affective factors that motivate users to use ChatGPT. The tool's novelty value has been found to be a driver of use in two studies (Ma & Huo, 2023; Wolf & Maier, 2024), and other studies have found that entertainment is correlated with the intention to continue using ChatGPT over time (Yin et al., 2023; Skjuve et al., 2023; Zhou & Zhang, 2024). Users are more likely to use ChatGPT when they trust the tool, and in the case of ChatGPT trust is reportedly based on the tool's interactivity and perceived sociality (Jung et al., 2024). The tool's anthropomorphism may be a motivator for some users (Pham et al., 2024), though, as mentioned below, it may drive away other users. Finally, some users may be motivated to use ChatGPT to avoid "falling behind" in a time of technological change (Yin et al., 2023).

The literature also suggests a constellation of detractors from ChatGPT use. Seemingly the greatest factor cutting against ChatGPT use is the accuracy of its output (Kim, Lee, et al., 2024; Kobiella et al., 2024; Wolf & Maier, 2024), coupled with the difficulty of controlling and customizing its output (Kim, Lee, et al., 2024; Kobiella et al., 2024; Skjuve et al., 2023; Sun et al., 2024). Anxiety is also a major theme in this literature; some users avoid ChatGPT because of technology anxiety (Li, 2025; Pham et al., 2024) or anxiety due to rapid change in the AI sector (Sun et al 2024). The tool's anthropomorphism and perceived human-like intelligence, which may be experienced as "creepy," can also detract from people's willingness to use ChatGPT (Baek & Kim 2023; Skjuve et al., 2023; Yin et al., 2023). Other reasons people may not use ChatGPT discerned in the literature include: a lack of feeling ownership over the tool's output (Kobiella et al., 2024; Wolf & Maier, 2024), privacy and security risks (Li, 2025; Wolf & Maier, 2024), limited trust in the tool because it doesn't cite sources accurately and reliably (Jung et al., 2024; Wolf & Maier, 2024), and the difficulty in some regions (e.g., China) finding resources to help with using ChatGPT (Sun et al., 2024).

Turning to what people are trying to accomplish by using ChatGPT—i.e., how the tool fits into people's broader activities, habits and information behaviors and practices—the literature is less detailed. Skjuve et al. (2024) reported on the categories of activities supported by ChatGPT, finding that ChatGPT is used for productivity, novelty, creative work, learning and development, entertainment, and social interaction and support. Productivity was the dominant category, comprising 55% of their data, which includes activities such as information retrieval (34% of all participants expressed using ChatGPT for information retrieval). In a similar study, Boryung and Brenton (2024) documented the types of usage early adopters made of ChatGPT, categorizing these use cases into: information seeking, ease and efficiency, speediness, entertainment, and social interaction. These findings are impressionistic and should be considered preliminary, seeing as they correspond to different notions of information need, overlap to some extent, and reflect different levels of abstraction (e.g., "information seeking" is about content, while "speediness" is about time).

Other studies report ChatGPT use cases incidentally or as illustrations in their findings: coming up with diverse ideas (Jung et al., 2024; Sun et al., 2024); polishing up some input (Sun et al, 2024); creating boilerplate for some communication (Jung et al., 2024). Relatedly, a study of user behavior with a Microsoft chatbot predating ChatGPT shows that conversation is a major use case for such tools, including for learning English, having a debate, or chatting about pop culture (Brinkman & Grudin, 2023). Taken together, these results suggest an opportunity for further research to systematize the use cases (final causes) of ChatGPT, particularly through the lens of information need.

**METHODS**

To identify the information needs and practices that ChatGPT supports, a qualitative content analysis was conducted on people's accounts of how they used ChatGPT. These accounts were collected from Reddit, one of the most popular sites on the internet. On Reddit, users gather in discussion forums called "subreddits" to discuss countless topics, usually anonymously.

For this study, I collected 205 public and anonymous vignettes of ChatGPT usage shared from 2023 to 2025 in which people described using ChatGPT for a specific purpose. Most of these vignettes (94%) came from the r/ChatGPT subreddit, where users discuss all manner of ChatGPT-related topics, and the remainder came from OpenAI's community forums (reached via link from the r/ChatGPT subreddit).

Data collection proceeded with Google searches restricted to the domain "reddit.com" with combinations of terms such as "chatgpt," "use case," "real-life task," etc. I only included in the dataset vignettes that were detailed enough to understand the particular way in which ChatGPT was used. For instance, a user saying they use ChatGPT simply "to learn about politics" was not included in the dataset, but one saying they use it for tracking the relationship between their eating and a disease was included.

I created a spreadsheet of the collected vignettes and conducted a qualitative content analysis to determine the characteristics of the dataset. This proceeded iteratively. In the end, three key columns describe the ways in which people use ChatGPT to address their information needs:

- **Domain**: What sphere of life the vignette involves (e.g., hobby, personal finance, health, job search)

- **ChatGPT Role**: What ChatGPT was used for in the interaction (e.g., summarizing text, writing code, translating)

- **Human Need**: What the person was trying to accomplish or satisfy (e.g., personal budget, learning Python, make a purchase decision)

In the final coding of the dataset, 25 different ChatGPT roles were identified. I then mapped these different roles using affinity diagramming to discern the relationships among them. Affinity diagramming is an interpretive process of physically arranging datapoints according to their conceptual relatedness (the analyst places datapoints that are conceptually closer, physically closer together), and then identifying and labeling clusters of tightly-related datapoints that emerge (Rogers et al., 2023). Like the initial coding, this was also an iterative process. This exercise resulted in six distinct categories of ChatGPT roles that fell along two axes; these roles represent information practices that ChatGPT supports.

**FINDINGS**

**Information Needs: Why People Use ChatGPT**

Based on the vignettes collected in this study, people come to ChatGPT with a broad range of information needs. These include information needs as that concept is traditionally understood (the need to receive information or become informed, whether to fill a knowledge gap or modulate a feeling), but they also demonstrate other types of information need, such as the need to create information, evaluate information, change information, etc., as well as situations where the information need seems to be a need for interaction rather than for the content of the information per se (e.g., a person who simply wants to talk). Here I discuss these needs as represented in the sample, first in terms of domain and then in terms of human need.

The results show that people use ChatGPT in a variety of domains, including work, home and family, leisure, and more. Of the 205 vignettes in the sample, 180 had a discernible domain. Of those, the plurality related to **home and family** (50, or 28%). These vignettes showed a broad range of tasks, including: writing party invitations, simple programming tasks (e.g., creating customized notifications for social media), understanding current events, getting media recommendations, talking about personal matters, seeing how well an outfit matches, tracking baby activities, creating a personal budget, and writing bedtime stories for one's children. Nearly as many vignettes related to related to **work** (48, or 27%). These vignettes included many forms of creating information, such as newsletter content, meeting materials, job descriptions, software documentation and customer service messages. Half of the "work" vignettes were of this nature. Other tasks represented in this domain included simple programming tasks (e.g., debugging a script), summarizing documents, making a plan (e.g., how to accomplish something in the most efficient way), and editing text (e.g., rewriting an email in a different tone). The next largest category was **leisure** (43, or 24%). All but 5 of these seemed to represent serious leisure (e.g., a hobby), generally centered on learning. For example, 9 vignettes report using ChatGPT to learn programming through practice and tutoring, and 4 report using it to study a new language through conversation, vocabulary quizzing and explanation of grammatical concepts. Other hobby-related tasks in the sample range from writing fan fiction, to preparing materials for a role-playing game (e.g., Dungeons & Dragons), to discussing philosophy or fiction, to helping a user beat a particular video game. The five vignettes representing casual leisure included one-off fun usages of ChatGPT, such as experimenting with the tool and using it to write a song for no particular purpose.

After these three major domains, there were five minority domains discerned in the sample: health, food, fitness, job search, and legal, each comprising 10 vignettes or fewer. The largest of these was the **health** domain (10, or 6%), which included tasks such as getting insight on one's medications and medical conditions, interpreting medical tests, generating questions to ask a doctor, getting therapy, and getting a skin care routine. Next, and related, were vignettes related to **fitness** (8, or 4%). These vignettes included tasks such as calorie tracking, meal planning and workout planning. Next was **job search** (8, or 4%), which included tasks such as writing or editing a cover letter and conducting a mock job interview. The next domain was **food** (7, or 4%), which included meal planning (unrelated to weight management) and providing recipes (e.g., based on ingredients at hand or with a particular substitution). Lastly, the smallest domain was **legal** (6, or 3%), which included tasks such as getting steps to change one's immigration status, summarizing a legal document and writing an official letter.

As implied in the discussion of domains above, people approached ChatGPT with a range of information needs. The largest category of these was **writing or editing** (59, or 29%), suggesting a person's need to create information. As mentioned above, these tasks ranged from writing emails for work or home life, writing fun texts for family, creating PowerPoint slides at work, writing job application materials and so on. The next largest category of need was **learning** (28, or 14%), either getting guided in how to do something or getting help solidifying one's knowledge. Following that was **simple programming tasks** (16, or 8%), whether for work or a hobby, such as debugging code, translating code to a different programming language, refactoring code (rewriting it for efficiency and/or clarity), creating a script for a "boring" function, etc. Another notable category not yet mentioned was **friend** (7, or 3%), representing the need for someone to vent to, confide in, and get advice from on personal issues.

Beyond these categories, a broad range of human needs are represented in the sample, most of which comprise 1% or less of the vignettes. To share a few examples, these include: having "book club"–type conversations, developing ideas for a new business, building a computer, making a purchase decision, planning a road trip, translating text (e.g., English to Spanish), overcoming writer's block, self-reflecting on personal events, and getting personal motivation. The next subsection discusses how people used ChatGPT concretely to address these needs.

**Information Practices: The Roles of ChatGPT**

As reported above, people came to ChatGPT with a spectrum of information needs. The next question is the role ChatGPT played in addressing those needs. There was not necessarily a one-to-one relationship between human needs and ChatGPT roles. For example, among the vignettes related to language learning, a subset represented the human need of better understanding a concept. In some of these vignettes, ChatGPT's role was answering questions in a Q&A back-and-forth way, while in other vignettes ChatGPT's role was offering lengthier explanations. To give another example, among the vignettes where the human need was to prepare for a job interview, sometimes ChatGPT's role was as a conversation partner (e.g., in a mock interview), while other times it was to generate divergent ideas (e.g., creating a list of possible interview questions). In all but 8 vignettes, a clear role of ChatGPT in the interaction was identified.

Orthogonal to the 75 different human needs identified in the dataset, 25 distinct roles for ChatGPT were discerned. These 25 roles were then mapped via affinity diagramming, resulting in six categories of ChatGPT roles organized along two axes: Writing, Deciding, Identifying, Ideating, Talking, and Critiquing (see Figure 1). These higher-level categories can be interpreted as information practices which ChatGPT supports.

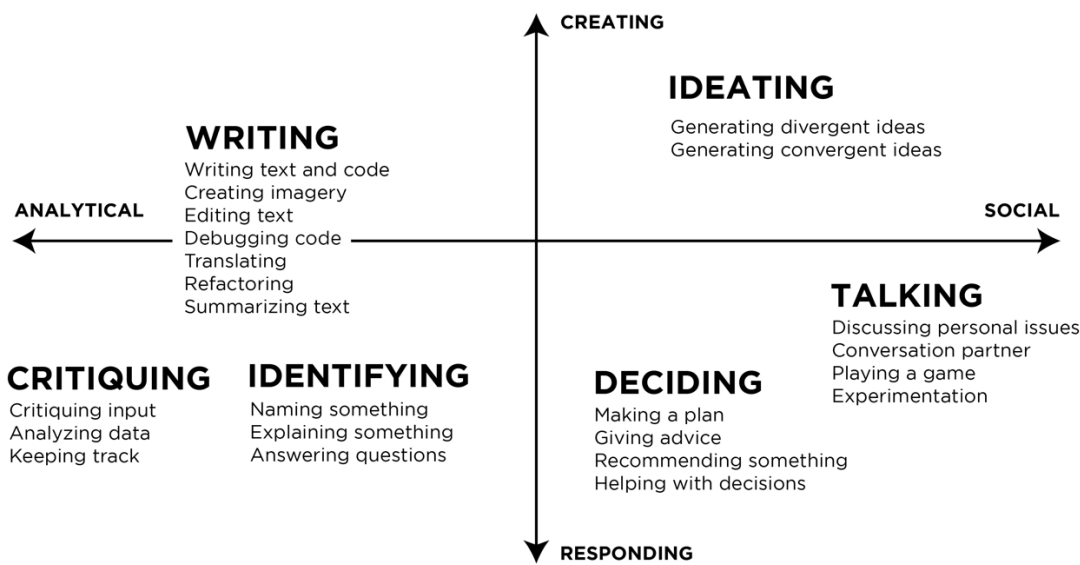

**Figure 1. Mapping of ChatGPT Roles**

*Writing*

Writing was the largest category, comprising 92 vignettes (47% of the sample). The most common ChatGPT role in this category was generating text, whether for public consumption (e.g., a social media post), for one-on-one communication (e.g., an email to a customer), or for oneself (e.g., motivational messages). This category also included writing code (e.g., a simple script or function) and generating imagery. (While ChatGPT is primarily a text-based tool, in late 2023 OpenAI made image generation and other multimodal features available within ChatGPT for paid users (OpenAI, 2023).)

Besides creating new material, this category also included various editing roles. For text, ChatGPT was used to paraphrase, modify tone, improve catchiness and fix grammar. It was also used to generate summaries of text. For code, ChatGPT was used to debug (find problems preventing code from working correctly) and refactor (rewrite code to be more efficient or readable while maintaining functionality). Finally, for both text and code, ChatGPT was used to translate—either between natural human languages or between programming languages.

*Deciding*

The next most prominent category was Deciding, comprising 42 vignettes (21% of the sample). This category involved helping people make decisions of various kinds, from recommending things (e.g., books, meal in a restaurant) and helping with purchasing decisions (e.g., comparing gym bags, finding the cheapest gasoline nearby), to giving advice (e.g., what stretches will help a certain neck ailment, how to deal with a social problem).

The largest subset of Deciding vignettes involved ChatGPT helping the user make a plan. There were a range of plan types in the sample, including: building a computer, customizing a recipe, fixing things around the house, getting out of debt, designing a curriculum, meal planning, making a shopping list, creating an outline for a book, and so on.

*Identifying*

Next, Identifying comprised 19 vignettes, or 10% of the sample. Most of these involved ChatGPT explaining something that the user requested—definitions, how something works, additional examples, etc. Relatedly, there were several examples of ChatGPT answering questions. In four vignettes, people reported using ChatGPT to name something based on a description, such as the name of a book based on a description of its plot and characters, or the name of a tool based on a photograph.

*Ideating*

Ideating, or generating ideas, was the next largest category of ChatGPT roles, with 17 vignettes (9% of the sample). For the most part, ChatGPT was used to develop divergent ideas—that is, to help the user explore new directions and concepts based on a starting point. In two cases, ChatGPT was used to develop convergent ideas, i.e., to generate new ideas within a more constrained conceptual space or based on a more solidified starting point. Incidentally, both these cases had to do with generating material for *Dungeons & Dragons* campaigns: helping the user ideate character backstories, monsters and scenarios for an encounter, etc. To be sure, in actual practice users likely go between moments of diverging and converging that were not fully captured in these vignettes—for instance, one user spoke of "bouncing around" ideas—so the distinction and ratio of divergent and convergent in this case should not be taken as rigid.

*Talking*

Perhaps the most obvious usage of ChatGPT, given its name, is talking. Still, Talking comprised only 8% of the sample, with 15 vignettes. The most common form of Talking in the sample was ChatGPT functioning as a conversation partner, whether for learning a new language, discussing some material, practicing a job interview or role-playing a life situation. Next most common was discussing a user's personal problems in the mode of what a human friend would do. Users wrote of "confiding" and "being understood," and of ChatGPT being "non-judgmental." Somewhat relatedly, one vignette involved ChatGPT being the user's partner in a two-player game, and another vignette involved having fun experimenting with ChatGPT's capabilities.

*Critiquing*

Finally, 12 vignettes (6% of the sample) were categorized as Critiquing. In this role, ChatGPT offered constructive feedback to a user's input. For example, people used ChatGPT to get constructive criticism on their writing or code, to check if their outfit matched, to describe the tone of messages the user wrote and to estimate the calories based on a photo of a plate of food. Also in this category was the role of keeping track; two people in the sample used ChatGPT to log the activities of an infant and provide a report at the end of the day, and others used it to track budgetary spending and caloric intake. Finally, ChatGPT was also used for data analysis, providing users with insights on their running data, laboratory test results and health symptoms.

**The Conceptual Space of ChatGPT Information Practices**

The analysis method of affinity diagramming not only helps identify categories but also the relationships among those categories. In this study, the mapping of ChatGPT roles revealed two axes on which the information practices supported by ChatGPT can be understood, as depicted in Figure 1.

The primary axis ranged from Analytical to Social. On this axis, Analytical refers to use cases that are primarily intellectual, cognitive, mechanistic and task-centered, while Social refers to use cases that are primarily relational, affective, open-ended and personal. Critiquing is the category furthest at the Analytical end, and Talking is furthest at the Social end. The other four categories range along this axis.

The secondary axis ranges from Creating to Responding. At the Creating end of the spectrum, the AI provides most of the content in the interaction, while at the Responding end, the user provides most of the content in the interaction. Writing is an example of Creating, and Identifying is an example of Responding.

**DISCUSSION**

As new technologies proliferate and evolve, information scholars must come to understand how these technologies are influencing and enabling people to engage with information—including how they need, seek, use and experience information. To date, the relevant research on generative AI focuses on users' motivations and user experience factors that lead to use or non-use of generative AI products. The present study sheds light on the information needs that people use generative AI tools such as ChatGPT to address, reporting on the domains and human needs of ChatGPT usage as well as the categories of information practice supported by ChatGPT in practice.

Beyond providing information about how a new technology is impacting information engagement, this study makes theoretical contributions to the field of information research more broadly. First, it demonstrates an approach to mapping the information practices relevant to a particular context, giving more theoretical clarity beyond the narrative descriptions and lists that are common in the literature.

Next, this study sheds new light on the concept of information need. As reviewed earlier, in the literature information need has been conceptualized as question answering, filling a knowledge gap or making sense. These conceptualizations of information need work well for cases of seeking information in the sense information is traditionally defined (e.g., retrieving a fact, locating a book or article). But today it is recognized that information literacy is not just about consuming information, but also about comparing information, manipulating information, creating information, and so on (ACRL, 2015). The notion of information need that ChatGPT responds to isn't just question-answering, filling a knowledge gap or making sense. It is also a more active need to "cope" with the world, to use a term from phenomenology, referring to the application of people's understanding and practical intelligence as they go through life (Dreyfus & Wrathall, 2014). This notion of information need can help us understand not just discrete moments of information seeking but also how such information seeking fits into a person's day and life. This perspective also suggests a bridge between traditional notions of information need and information behavior to more holistic framings such as information experience.

**Opportunities for Further Research**

This was a qualitative study conducted by a single researcher, and as such it is limited in its generalizability. However, it provides a clear and decisive contribution that can serve as groundwork for further research. This section reflects on the nature of this study and opportunities for future work.

First, this study involved data collected from Reddit, a platform that is U.S.-centric and primarily Anglophone. Further work could use other data collection methods to offer a more global perspective of generative AI usage. Though, to be sure, this same limitation also applies to generative AI products themselves (Tuna et al., 2024).

Another limitation of using Reddit for data collection is that this study could only include use cases that people shared publicly. While Reddit's anonymous and ecumenical nature enables users to share things they may keep private in other contexts, there are likely things that people may keep private even from Reddit—or perhaps that they are not even aware of. The findings of this study could thus be triangulated with those using other data gathering methods.

This study was also limited to ChatGPT, a single generative AI tool, albeit the most widely used one. Further research should examine how people use other generative AI products (e.g., Claude, Gemini, Llama, etc.) to address their information needs. Anecdotally, people do report using different products for different purposes; and emerging research suggests that different generative AI products have different "personalities," which may lead to different use cases (Serapio-García et al., 2025).

Next, this study provided a single snapshot of how ChatGPT is being used, collapsing use cases from a two-year period. Generative AI tools, like any technology, evolve over time, and it is likely that the use cases supported by these products also evolve. Further work examining this chronological evolution may reveal additional insights about people's information engagement with emerging technologies.

Another path for future work would be to examine ChatGPT as an information source through a process model of information behavior. This study sought to categorize the different types of information use cases supported by ChatGPT, but it would be also be useful to examine any or all of these use cases temporally. Such research could, for example, reveal whether the way people interact with ChatGPT aligns with existing process models in information behavior, such as those reviewed earlier in this paper.

Finally, this work suggests further opportunities to learn how ChatGPT and other such tools address (or not) people's information needs successfully. Are there certain use cases for which generative AI tools are more satisfying or less satisfying? Why do people choose to go to a generative AI tool for a given information need versus a different information source, such as a person or a webpage?

**CONCLUSION**

This study framed ChatGPT as an information source, investigating the information needs that people come to ChatGPT with and the information practices that ChatGPT supports. The findings show that ChatGPT is used in a range of life domains (home/family, work, leisure, etc.) and for a range of human needs (writing/editing, learning, simple programming tasks, etc.), constituting the information needs that people use ChatGPT to address. Related to these information needs, the findings show six categories of information practices that ChatGPT supports: Writing, Deciding, Identifying, Ideating, Talking, and Critiquing. This work suggests that, in the AI age, information need should be conceptualized not just as a matter of "getting questions answered" or even "making sense," but as skillfully coping in the world, a notion that includes both understanding and action. As discussed, this study leads to numerous opportunities for future work at the junction of generative AI and information needs, seeking, use and experience.

**GENERATIVE AI USE**

We confirm that we did not use generative AI tools/services to author this submission.